\documentclass[a4paper,twoside,11pt]{article}
\usepackage[a4paper,bindingoffset=0.2in,
left=1in,right=1in,top=1in,bottom=1in,
footskip=.25in]{geometry}
\usepackage{amsmath}
\numberwithin{equation}{section}
\usepackage{amsfonts}
\usepackage{amssymb}
\usepackage{graphicx}
\usepackage{hyperref}
\usepackage{fancyhdr} %for header and footer
\usepackage{xcolor}
\begin{document}
	\title{Heun polynomials and exact solutions for the massless Dirac particle in the   $C$-metric}
	\author{Priyasri Kar$^a$, Ritesh K. Singh$^b$, Ananda Dasgupta$^c$, 
		Prasanta K. Panigrahi$^d$\\[0.2cm]
		{\it Department of Physical Sciences,}\\ {\it Indian Institute of Science Education and Research Kolkata, Mohanpur 741246, India}\\[0.2cm]
		$^a$ {\tt pk12rs063@iiserkol.ac.in},
		$^b$ {\tt ritesh.singh@iiserkol.ac.in},\\
		$^c$ {\tt adg@iiserkol.ac.in},
		$^d$ {\tt pprasanta@iiserkol.ac.in}
	}
	\maketitle
	\begin{abstract}
		The equation of motion of a massless Dirac particle in the $C$-metric leads to the general Heun equation (GHE) for the radial and the polar variables. The GHE, under certain parametric conditions, has been cast in terms of a new set of $su(1,1)$ generators involving differential operators of \emph{degrees} $\pm 1/2$ and $0$. Additional \emph{Heun polynomials} are obtained using this new algebraic structure and are used to construct some exact solutions for the radial and the polar parts of the Dirac equation.
		
	\end{abstract}
	
	\section{Introduction}\label{intro}
	The $C$-metric was first discovered by Levi-Civita~\cite{levicivita} and has obtained its name and place in the celebrated classification of exact space-times by Ehlers and Kundt~\cite{ehlers_kundt}. It represents the exterior gravitational field of a uniformly accelerating Schwarzschild black hole under certain conditions~\cite{exact_spacetimes_book}. The line element of the $C$-metric in spherical like co-ordinates is given by
	\begin{equation}\label{$C$-metric}
	ds^2=\dfrac{1}{\Omega^2(r,\theta)}\left(\dfrac{Q(r)}{r^2}dt^2-\dfrac{r^2}{Q(r)}dr^2-\dfrac{r^2}{P(\theta)}d \theta^2-r^2P(\theta) \mbox{sin}^2 \ \theta d \phi^2 \right),
	\end{equation}
	where,
	\begin{equation}\label{metric_functions}
	Q(r)=r\left(r-2M\right)\left(1-A^2r^2\right), \qquad P(\theta)=1-2MA \ \mbox{cos}\theta, \qquad \Omega(r,\theta)=1-Ar \ \mbox{cos} \theta
	\end{equation}
	and the constants $M(\ge0)$ and $A(\ge0)$ denote the mass and acceleration of the source, respectively.
	The metric reduces to that of a Schwarzschild black hole for $A=0$ and to the Rindler metric representing flat spacetime in uniformly accelerating co-ordinates for $M=0$.  The behaviour of massless spin-1/2 particles in this metric has been studied in Ref.~\cite{bini_bit_ger}, where
	the radial and polar parts of the Dirac equation turn out to be general Heun equations (GHE) upto some variable transformations.
	
	The GHE is the most general form of second order Fuchsian differential equations with four regular singularities. Due to its appearence in a wide variety of physical problems~\cite{hortacsu_main}, the properties and solutions of this equation have been studied extensively~(\cite{ronveauxbook, slavyanov_lay, maier}) in the past few decades. Our present target is to discuss some of the exact solutions of this equation using the algebraic properties of the general Heun operator and use them to study the dynamics of massless Dirac particle in the   $C$-metric. Algebraic methods have played significant role in unravelling the symmetry of second order differential operators~\cite{willard_miller,vilenkin}. A partial algebraization of the whole solution space has been achieved~\cite{turbiner,shif_tur,shifman,ush_book} for a wide
	class of quantal quasi-exactly solvable (QES) Hamiltonians in one or more dimensions. Noticeably, the study of these QES systems boils down to the study of the algebraic properties of the Heun operator~\cite{turbiner_big}. More specifically, a second order differential equation is quasi-exactly solvable if a corresponding Heun operator can be cast in terms of a set of linear differential operators, that constitute a finite dimensional representation of the $sl(2)$ algebra.
	
	Inspired by this algebraic approach, we here present a different two parametric representation of $su(1,1)$ generators of \emph{degrees}\footnote{The degree $d$, of an operator $O_d$, is defined as the change in the power of a monomial, when acted upon by it, i.e., $O_d \ z^p \propto z^{p+d}$. } 
	$\pm1/2$ and $0$ in contrast to the set of integer degree generators involving a single parameter, existent in the literature~\cite{turbiner}. The necessary parametric conditions for casting the GHE in terms of these new generators require the singularities at zero and infinity of the equation to be \emph{elementary}\footnote{Any regular singularity can be characterized by two exponents $\rho_1$ and $\rho_2$, which are the two roots of the indicial equation. The cases with $|\rho_1-\rho_2|=1/2$ are known as elementary singularity and are of special significance. All regular and irregular singularities can be obtained by the coalescence of two and three or more elementary singularities, respectively~\cite{ince}.}.
	In the case of massless Dirac particle in the   $C$-metric, the singularity at 0 of the radial Heun equation and that at $\infty$ of the polar Heun equation are observed to satisfy this condition irrespective of the parameter values. The other required elementary singularity for each of the equation is ensured by imposing necessary parametric conditions. The advantage of using these new generators is that it tracks the quasi-polynomial solutions (known as \emph{Heun polynomials}~\cite{ronveauxbook}) admitted by the GHE, besides the polynomial solutions that are found using the existent generators~\cite{turbiner}. The Heun polynomials, being closed form solutions that are valid in the entire complex plane (except for the singular points and possibly with appropriate cuts to ensure single-valuedness), are of immense physical importance and thus have been explored extensively~\cite{poly_condns}.
	In this work we apply our method to obtain Heun polynomials and use them to construct exact (closed form) solutions for the radial and polar parts of the Dirac equation for some specific values of the separation constant.

	The organisation of the paper is as follows: In section \ref{casting_&_rep_theory} we cast the GHE in terms of the newly introduced $su(1,1)$ generators and briefly discuss Heun polynomials in connection with the finite dimensional representations of $su(1,1)$. In section~\ref{solutions} we explicitly construct the most general forms of some of the Heun polynomials, available from the present method. Section \ref{example} deals with the dynamics of the massless Dirac particle in the   $C$-metric, where we read off the solutions from section \ref{solutions} for the radial and polar parts of the Dirac equation, using appropriate values of the parameters. Finally in section \ref{conclusions}, we summarise our work and conclude with some future visions.

	\section{Recasting of the Heun operator and the $su(1,1)$ representations}\label{casting_&_rep_theory}
	The Heun equation, in its canonical form, is given by~\cite{ronveauxbook}
	\begin{equation}\label{heuneqn}
	\frac{d^2y(z)}{dz^2}+\left(\frac{\gamma}{z}+\frac{\delta}{z-1}+\frac{\varepsilon}{z-a}\right)\frac{dy(z)}{dz}+\frac{\alpha\beta
		z-q}{z(z-1)(z-a)}y(z)=0,
	\end{equation}
	with regular singularities at $z$ = $0, \ 1, \ a(\ne 0,1) \mbox{ and} \ \infty$.
	The exponents at these singularities are $(0, \ 1-\gamma)$, $(0,\ 1-\delta)$, 
	$(0, \ 1-\varepsilon)$ and $(\alpha, \ \beta)$, respectively. Here $q$ plays 
	the role of the \emph{eigen parameter}. In this work we study Heun equations with 
	only real parameters. The equation, being a second order linear 
	homogeneous equation with $4$ regular singularities, satisfies the Fuchsian condition: 
	\begin{equation}\label{fuchsiancondn}
	\gamma+\delta+\varepsilon=\alpha+\beta+1,
	\end{equation}
	allowing elimination of $\varepsilon$ in favor of the others. 
	Eq.~(\ref{heuneqn}) can be written as:
	\begin{equation}\label{gendiffeqn}
	\mathcal{H}y(z)\equiv\left[f_{1}(z)\frac{d^2}{dz^2}+f_2(z)\frac{d}{dz}+f_3(z)\right]y(z)
	= 0,
	\end{equation}where $f_{1}(z)=a_{0}z^3+a_{1}z^2+a_{2}z$, $\space$ $f_{2}(z)=a_{3}z^2+a_{4}z+a_{5}$ $\space$ and $\space$ $f_{3}(z)=a_{6}z+a_{7}$.
	%\endnumparts
	The parameters $a_{i} \in \mathbb{R}, \ \mbox{for} \ i = 0,\ldots,7 $ are 
	given by
	\begin{subequations}\label{a0toa7}
		\begin{align}
		a_0 =& 1 \qquad \qquad \qquad a_3 = 1+\alpha+\beta \qquad \qquad \qquad \qquad
		\qquad a_6 = \alpha\beta \label{a0a3a6} \\
		a_1 =& -(a+1) \hspace{.9cm} a_4 = -[a\gamma+a\delta-\delta+\alpha+\beta+1]
		\hspace{.9cm} a_7 = -q \label{a1a4a7}\\
		a_2 =& a \qquad \qquad \qquad a_5 =  a\gamma. \label{a2a5} 
		\end{align}
	\end{subequations}
	Evidently, the Heun equation consists of differential operators of 
	 degrees $+1,\ 0$ and $-1$. Denoting them by $P_+$, $F(P_0)$ and $P_-$ respectively, Eq.~(\ref{gendiffeqn}) can be rewritten 
	as~\cite{Arunesh}
	\begin{equation}
	\mathcal{H}y(z)=[P_+ + F(P_0) + P_-] y(z) = 0 
	\label{heunclass}
	\end{equation}
	where,
	\begin{eqnarray}
	P_{+} = a_{0}z^3\frac{d^2}{dz^2}+a_{3}z^2\frac{d}{dz}+a_{6}z, \quad 
	P_{0} = z\frac{d}{dz}-j,\quad 
	P_{-}=a_{2}z\frac{d^2}{dz^2}+a_{5}\frac{d}{dz} \\ 
	\mbox{and} \quad F(P_0)=a_{1}P_0^2+\left((2j-1)a_1+a_4\right)P_0
	+\left(j(j-1)a_1+ja_4+a_7\right).
	\end{eqnarray}
	The above operators satisfy cubic deformation of $sl(2)$ algebra among themselves, which has been exploited~\cite{Arunesh} to track a part of the eigenspace of a Heun operator analytically. However, non-linear deformations of $sl(2)$ are associated with non-trivial representation theory~\cite{rocek}. Our present purpose is to cast the Heun opeartor in terms of the new two parameter $su(1,1)$ generators and for the convenience of this casting, we use the above degreewise classification of operators (i.e., $P_+$, $P_-$ and $F(P_0)$), as shown below. 
		
	\subsection{The new $su(1,1)$ generators and the Heun operator}\label{casting}
	With the aim to cast the Heun operator in terms of linear $su(1,1)$ algebra, we propose the following representation of $su(1,1)$ generators
	\begin{eqnarray}\label{boperators}
	E_+ = 2 z^{3/2}\frac{d}{dz}-2\mu\sqrt{z} \label{bplus}, \quad
	H = 2 z\frac{d}{dz}-\left(\mu+\nu\right) \label{bzero} \quad \mbox{and} \quad
	E_- = 2 z^{1/2}\frac{d}{dz}-\frac{2\nu}{\sqrt{z}}, \label{bminus}
	\end{eqnarray}
	where the two parameters $\mu$ and $\nu$ are to be determined from suitable inputs. The operators $E_+$,
	$H$ and $E_-$ have degrees $+1/2, \ 0$ and $-1/2$, respectively. They satisfy $su(1,1)$ algebra in the monomial space, with
	commutators,
	\begin{eqnarray}\label{bcommutators}
	[H,E_{\pm}]=\pm E_{\pm}, \hspace{1cm} [E_+,E_-]=-2H
	\end{eqnarray}
	and the Casimir:
	\begin{equation}\label{unshifted_casimir}
	C(\mu,\nu)=\dfrac{1}{2}\left(E_+E_-+E_-E_+\right)-H^2=-(\mu-\nu)(\mu-\nu+1).
	\end{equation}
	
	We now proceed to construct the differential operators $P_+, P_-$ and $F(P_0)$ from the generators~(\ref{boperators}) and identify the conditions under 
	which this is possible. The generators are of degrees $\pm1/2$ and 
	$0$, whereas the degrees of operators $P_{\pm}$ and $F(P_0)$ are $\pm1$ and 
	$0$, respectively. Hence, it is clear that the complete Heun operator will 
	comprise of linear and quadratic forms of the generators.
	Assuming the forms
	\begin{equation}
	P_+ = c_+ E_+E_+ \qquad \mbox{and}  \qquad P_- = c_- E_-E_-
	\end{equation}
	where $c_{\pm}$ are constants, we obtain
	\begin{eqnarray}
	&& a_0 = 4c_+, \ \ \ 
	a_3 = \dfrac{a_0}{2}(3-4\mu),  \ \ 
	a_6 = \dfrac{a_0}{2}\mu(2\mu-1),\label{a0a4a7}\\
	&& a_2 = 4c_-, 
	\quad a_5 = \dfrac{a_2}{2}(1-4\nu), \quad
	\dfrac{a_2}{2}\nu(2\nu+1) = 0.\label{a6a2}
	\end{eqnarray}
	Using Eqs.~(\ref{a0a4a7}) and (\ref{a0a3a6}) one obtains:
	\begin{equation}\label{factor_condn1}
	|\alpha-\beta|=\dfrac{1}{2},
	\end{equation}
	implying the singularity at $z=\infty$ is elementary.
	Similarly, using Eqs.~(\ref{a6a2}) and (\ref{a2a5}) one is left with 
	just two values of $\nu$:
	\begin{equation}\label{factor_condn2}
	\nu=0 \implies a_5/a_2=\gamma=1/2 \qquad \mbox{or,} \qquad \nu=-1/2 \implies 
	a_5/a_2=\gamma=3/2.
	\end{equation}
	Above solutions of $\gamma$ imply that the singularity at $z=0$ is also 
	elementary.
	In terms of the equation parameters, $\mu$ and $\nu$ are given as:
	\begin{equation}\label{mu_nu_form}
	\mu=\mbox{max}\{-\alpha,-\beta\} \quad \mbox{and} \quad \nu=\mbox{min}\{0,1-\gamma\}.
	\end{equation}
	Finally, $F(P_0)$ can be written in the form 
	\begin{equation}
	F(P_0) = c_2HH+c_1H+c_0,
	\end{equation}
	where,
	\begin{eqnarray}\label{a1a5a8}
	\nonumber a_1 = 4c_2, \quad 
	a_4 = 2(c_1-2c_2(\mu+\nu-1)) \\ 
	\mbox{and} \quad a_7 =-q=
	c_0+(\mu+\nu)(c_2(\mu+\nu)-c_1).
	\end{eqnarray}
	Thus, a Heun operator with elementary singularities at $z=0$ and $z=\infty$,
	i.e., with parametric conditions Eqs.~(\ref{factor_condn1}) and
	(\ref{factor_condn2}), becomes
	\begin{equation}\label{factorized_Heun}
	\tilde{\mathcal{H}}\equiv c_+E_+E_+ + c_-E_-E_- + c_2HH + c_1H + c_0,
	\end{equation}
	where we use the tilde to distinguish it from any arbitrary Heun operator. To write $\tilde{\mathcal{H}}$ in differential form we choose $\beta=\alpha+1/2$, implying
	\begin{eqnarray}
	\mu=-\alpha\label{mu_alpha_reln}
	\end{eqnarray}
	and obtain
	\begin{equation}\label{heun4degreehalf}
	\frac{d^2y(z)}{dz^2}
	+\left(\frac{\gamma_{i\{i=1,2\}}}{z}+\frac{\delta}{z-1}+\frac{\varepsilon}{z-a}\right)\frac{dy(z)}{dz}+\frac{\alpha\left(\alpha+\frac{1}{2}\right)
		z-q}{z(z-1)(z-a)}y(z)=0,
	\end{equation}
	where $\gamma_1=1/2$ and $\gamma_2=3/2$. The Fuchsian condition~(\ref{fuchsiancondn}) now reads
	\begin{equation}\label{constrainedFuchcondn}
	\gamma_i+\delta+\epsilon=2\alpha+3/2.
	\end{equation}
	From here onwards we shall deal with Heun equations of the form Eq.~(\ref{heun4degreehalf}). With the Heun operator $\tilde{\mathcal{H}}$ casted as Eq.~(\ref{factorized_Heun}), one can use the representations of $su(1,1)$ to look for solutions, as discussed below.
	
	\subsection{Finite dimensional representations of $su(1,1)$ and Heun polynomials} \label{rep_theory_and_heun_sols}
	In this section we discuss Heun polynomials in connection with the finite dimensional representations of $su(1,1)$. Recasting the Heun operator in terms of the $su(1,1)$ generators $E_+$, $E_-$ and $H$ solves for $\mu$ and $\nu$ (Eqs.~(\ref{a0a4a7}) and(\ref{a6a2})). For non-negative half-integer values (including half-even integers, i.e., the integers) of $(\mu-\nu)=j \ (\mbox{say})$, there exist finite $(2j+1)$-dimensional representation spaces of $su(1,1)$. The states $|j,h\rangle$ of these representations are labelled by $j$ and the eigenvalue $h$ of $H$, with the range of $h$ given by $h \in \left[-j,j\right]$. Due to the quadratic dependence of $\tilde{{\mathcal H}}$ on $E_+$ and $E_-$, we note that each representation space splits into two subspaces as $V_j = V_e \oplus V_o$, consisting of the alternate states of the representation space, both of which are invariant with regard to the action of $\tilde{{\mathcal H}}$. As a result, a pair of solution spaces is obtained from every representation space. The subscripts $'e'$ and $'o'$ stand for even and odd respectively, odd subspace being the one containing the lowest weight (lowest $h$ value). The representation space is the space of monomials, which means that the state $|j,h\rangle \propto z^p$, where the power of the monomial is given by
	\begin{equation}\label{monomial_power}
	p=(h+\mu+\nu)/2.
	\end{equation}
	In other words, the set of $(2j+1)$ monomials constitute a $(2j+1)$-dimensional invariant subspace of the generators~(\ref{boperators}) and thus of the Heun operator~(\ref{factorized_Heun}). Clearly, having non-negative half-integer $(\mu-\nu)$ is a necessary (not sufficient) condition for the Heun equation to admit a quasi-polynomial solution, which is a linear combination of a finite number of monomials\footnote{For other values of $(\mu-\nu)$, there exists power series solutions of the Heun equation, the algebraic properties of which will be explored in a future work~\cite{my_next_work}.}.
	In terms of the equation parameters, this condition reads:
	\begin{subequations}\label{finite_dim_condns}
		\begin{align}
		-\alpha=& j \quad \mbox{when} \quad \gamma=1/2 \label{finite_dim_condn1} \\
		\mbox{or,} \quad -\alpha+1/2=& j \quad \mbox{when} \quad \gamma=3/2. \label{finite_dim_condn2}
		\end{align}
	\end{subequations}
	Once condition~(\ref{finite_dim_condns}) is satisfied, the Heun operator~(\ref{factorized_Heun}) (without the eigenparameter $q$) is made to act upon the finite dimensional monomial space. Finding the quasi-polynomial solutions then reduces to standard ($2j+1$)-dimensional matrix eigenvalue problem. Thus $(2j+1)$ linearly independent eigensolutions are obtained, each with a specific value of the eigenparameter $q$. These are Heun polynomials of the form $z^t P_N(z)$ (with $t \in \mathbb{R}$), where $P_N(z)$ is any arbitrary polynomial in $z$ of degree $N$. Among them, the polynomial solutions corresponding to $t=0$ can be obtained using the $su(1,1)$ generators in Ref.~\cite{turbiner}. The main purpose of the present work, besides providing a new algebraic structure of Heun equation, is to detect all the Heun polynomials of the form $z^t P_N(z)$ admtitted by the GHE in concern.

	\section{The Heun polynomials}\label{solutions}
	In this section we explicitly show the method of obtaining Heun polynomials from the finite dimensional representation spaces of $su(1,1)$. We provide, for the two allowed values of $\gamma$ (Eq.~(\ref{factor_condn2})), the most general forms of some of the Heun polynomials (upto triplet) for the convenience of use in an actual physical problem.
	
	\subsection{Case-1: {$ \gamma=1/2$}} \label{gamma_half} For this case we have $\nu=0$  from Eq.~(\ref{factor_condn2}) and the Heun equation (Eq.~(\ref{heun4degreehalf})) satisfies Fuchsian condition $\delta+\epsilon=2\alpha+1$.
	To obtain Heun polynomials from the finite dimensional representation spaces of $su(1,1)$, Eq.~(\ref{finite_dim_condn1}) implies that we must have
	\begin{equation}\label{gamma_half_alpha_values}
	\alpha \in \{0, -1/2, -1, -3/2, \dots\}.
	\end{equation}
	With the $\alpha$ value belonging to this set and solving for the appropriate value of the eigenparameter $q$, classes of Heun polynomial solutions for this case are obtained below.
	
	{\bf Singlet:} $\alpha=0$ ($\mu=0$, Eq.~(\ref{mu_alpha_reln})). The representation space consists of a single state with $H$ eigenvalue $h=0$. The power of the corresponding monomial is found using Eq.~(\ref{monomial_power}). Thus, we have a single solution:
	\begin{equation}\label{case1_singlet_soln}
	y=\mbox{constant} \quad \mbox{with eigenvalue} \quad q=0
	\end{equation}
	
	{\bf Doublet:} $\alpha=-1/2$ $(\mu=1/2)$. The solution space splits into the even and the odd spaces, as expected. The space $V_o$ consists of a single state with $h=-1/2$ and $V_e$ contains the state $h=1/2$. The corresponding monomials powers are 0 and 1. The two linearly independent solutions are:
	\begin{eqnarray}
	y_1 \in V_o&=& \mbox{constant}, \quad \mbox{with eigenvalue} \quad q=0 \label{case1_doublet_soln1} \\
	y_2 \in V_e&=& \sqrt{z}, \quad \mbox{with eigenvalue} \quad q=\dfrac{1}{2}(1-a)\delta, \label{case1_doublet_soln2}
	\end{eqnarray}
	
	{\bf Triplet:} $\alpha=-1$ $(\mu=1)$. The solution space splits as usual and we obtain three linearly independent solutions, two from the doublet($V_o$) subspace with $h$ eigenvalues \{-1,1\} and one from the singlet($V_e$) subspace with $h$ eigenvalue 0. The solutions are:
	\begin{eqnarray}
	\nonumber y_1 \in V_o&=& \frac{1}{2} \left\{ (a-1) (1+2 \delta)+\sqrt{4a+(a-1)^2(1+2\delta)^2}\right\}+z, \\ \mbox{with eigenvalue} \quad q&=&\frac{1}{4} \left\{-(a-1) (1+2 \delta) + \sqrt{4a+(a-1)^2(1+2\delta)^2}\right\} \label{case1_triplet_soln1} \\
	\nonumber y_2 \in V_o&=& \frac{1}{2} \left\{ (a-1) (1+2 \delta)-\sqrt{4a+(a-1)^2(1+2\delta)^2}\right\}+z, \\ \mbox{with eigenvalue} \quad q&=&\frac{1}{4} \left\{-(a-1) (1+2 \delta) - \sqrt{4a+(a-1)^2(1+2\delta)^2}\right\} \label{case1_triplet_soln2}\\
	\mbox{and} \quad y_3 \in V_e&=& \sqrt{z}, \quad \mbox{with eigenvalue} \quad q=\frac{1}{2} (1+\delta-a \delta), \label{case1_triplet_soln3}
	\end{eqnarray}
	Among the solutions listed above, the ones given by Eqs.~(\ref{case1_singlet_soln}), (\ref{case1_doublet_soln1}), (\ref{case1_triplet_soln1}) and (\ref{case1_triplet_soln2}) are obtained from the existent $su(1,1)$ structure~\cite{turbiner} of the equation. However, the additional Heun polynomials involving $\sqrt{z}$, given by Eqs.~{\ref{case1_doublet_soln2}} and (\ref{case1_triplet_soln3}), are obtained using the new $su(1,1)$ structure presented here. This shows the advantage of using generators of degrees $\pm$1/2 and 0 in the present work.
		
	\subsection{Case-2: {$ \gamma=3/2$}} \label{gamma_three_half}
	Here we have $\nu=-1/2$ from Eq.~(\ref{factor_condn2}),
	with the Fuchsian condition reducing to $\delta+\epsilon=2\alpha$.
	Now, the necessary condition (Eq.~(\ref{finite_dim_condn2})) to find Heun polynomials from the finite dimensional representation spaces is
	\begin{equation}\label{gamma_three_half_alpha_values}
	\alpha \in \{1/2, 0, -1/2, -1, \dots\}.
	\end{equation}
	For $\alpha$ belonging to the above set and with appropriate value of $q$, some of the Heun polynomials in this case are listed below.
	
	{\bf Singlet:} $\alpha=1/2$ ($\mu=-1/2$, Eq.~(\ref{mu_alpha_reln})). There is a single state in this representation space with $H$ eigenvalue $h=0$. The corresponding monomial power is obtained from Eq.~(\ref{monomial_power}). Thus, the singlet solution is
	\begin{equation}\label{case2_singlet_soln}
	y=\dfrac{1}{\sqrt{z}}, \quad \mbox{with eigenvalue} \quad q=\frac{1}{2} (1+(a-1) \delta)
	\end{equation}
	
	{\bf Doublet:} $\alpha=0$ $(\mu=0)$. The doublet space, as earlier, splits into the even and the odd subspaces, which are two singlets. The powers of the monomials in $V_o$ and $V_e$ are $-1/2$ and $0$(constant), respectively. The two linearly independent solutions are:
	\begin{eqnarray}
	y_1 \in V_o&=& \dfrac{1}{\sqrt{z}}, \quad \mbox{with eigenvalue} \quad q=\frac{1}{2} (a-1) \delta \label{case2_doublet_soln1} \\
	y_2 \in V_e&=& \mbox{constant}, \quad \mbox{with eigenvalue} \quad q=0, \label{case2_doublet_soln2}
	\end{eqnarray}
	
	{\bf Triplet:} $\alpha=-1/2$ $(\mu=1/2)$. The solution space splits into a doublet($V_o$) consisting two monomials of powers $-1/2$ and $1/2$ and a singlet($V_e$) containing a monomial of power $0$. The three linearly independent solutions are
	\begin{eqnarray}
	\nonumber y_1 \in V_o&=& \dfrac{\frac{1}{2} (a-1) (1+2 \delta) +\sqrt{4a+(a-1)^2(1+2\delta)^2}}{\sqrt{z}}+\sqrt{z}, \\ \mbox{with eigenvalue} \quad q&=&\frac{1}{4} \left(-1-a +\sqrt{4a+(a-1)^2(1+2\delta)^2} \right) \label{case2_triplet_soln1}, \\
	\nonumber y_1 \in V_o&=& \dfrac{\frac{1}{2} (a-1) (1+2 \delta) -\sqrt{4a+(a-1)^2(1+2\delta)^2}}{\sqrt{z}} +\sqrt{z}, \\ \mbox{with eigenvalue} \quad q&=&\frac{1}{4} \left(-1-a -\sqrt{4a+(a-1)^2(1+2\delta)^2} \right) \label{case2_triplet_soln2}\\
	\mbox{and}\quad y_3 \in V_e&=& \mbox{constant}, \quad \mbox{with eigenvalue} \quad q=0. \label{case2_triplet_soln3}
	\end{eqnarray}
	Among the above solutions of Eq.~(\ref{heun4degreehalf}) with $\gamma=3/2$, given by Eqs.~(\ref{case2_singlet_soln}) to (\ref{case2_triplet_soln3}), the ones given by Eqs.~(\ref{case2_doublet_soln2})
	and (\ref{case2_triplet_soln3}) may be obtained from the methods of Ref.~\cite{turbiner}, while the Heun polynomials given by Eqs.~(\ref{case2_singlet_soln}), (\ref{case2_doublet_soln1}), (\ref{case2_triplet_soln1}) and (\ref{case2_triplet_soln2}) are additional ones obtained using the new $su(1,1)$ structure of the equation.

	With all the above Heun polynomials ready at our disposal, different physical problems involving the GHE can be addressed and the solutions can be read off with suitable parameter values. This is demonstrated in the next section, where we construct some exact solutions for the massless Dirac particle in the   $C$-metric.
	
	\section{The massless Dirac particle in the   $C$-metric}\label{example}
	In this section we study the dynamics of the massless Dirac particles in the   $C$-metric, the line element of which is given by Eqs.~\ref{$C$-metric} and \ref{metric_functions}. As mentioned earlier, this metric reduces to the Schwarzschild metric and to the Rindler metric in the two extreme limits $A=0$ and $M=0$, respectively. For the preservation of metric signature one requires $P(\theta)>0$, i.e., $\mbox{cos} \ \theta<1/2MA$ and to ensure it for all $\theta \in \left[0, \pi \right]$, $MA<1/2$ and $\phi \in \left[0,2 \pi/P(0)\right]$. The quantity $2MA(\equiv \eta)$ is the acceleration parameter of the source. The physical region of interest for the solution of the radial equation is $2M<r<1/A$, i.e., between the Schwarzschild horizon $r=2M$ and the Rindler horizon $r=1/A$. This requires $2M<1/A$, which implies $\eta<1$.
	
	The Dirac equation is obtained using NP formalism in the Kinnersley type null frame~\cite{kinnersley}. The wave function associated with the Dirac particles is written in terms of a pair of spinors. Due to the existence of the time-like and rotational Killing vectors $\partial_t$ and $\partial_\phi$ in the $C$-metric, the temporal and azimuthal parts of the wave function have the conventional form  $e^{-\iota(\omega t-m \phi)}$. On separation of variables, the radial and the polar parts of the Dirac equation turn out to be of the general Heun form up to some variable transformations. Below we discuss the solutions of these GHEs one by one.
	
	\subsection{The radial equation}\label{radial}
	The radial equation is given by~\cite{bini_bit_ger}
	\begin{equation}\label{radial_eqn}
	Q(r)^{-s}\dfrac{d}{dr}\left(Q(r)^{s+1}\dfrac{d {}_{s}\mathcal{R}\left(r\right)}{dr}\right) + V_{(rad)}(r){}_{s}\mathcal{R}\left(r\right)=0
	\end{equation}
	where,
	\begin{eqnarray}\label{radial_pot}
	\nonumber V_{(rad)}(r)&=&-2rA^2(r-M)(1+s)(1+2s)+ \dfrac{\omega^2r^4}{Q(r)} \\
	&&-2 \iota s \omega r\left(\dfrac{M}{r-2M}-\dfrac{1}{1-A^2 r^2}\right) -\lambda^2 +s\left(1+2s\right).
	\end{eqnarray}
	The label $s$ in ${}_{s}\mathcal{R}\left(r\right)$ can assume the values $\pm 1/2$ corresponding to the two radial functions and $\lambda$ is the separation constant. Effecting the transformation 
	\begin{equation}\label{radial_rescaling}
	{}_s\mathcal{R}(r)=u(z)Q(r)^{-s}\left(r-2M\right)^{k_1} \left(1+Ar\right)^{k_2}\left(Ar-1\right)^{k_3}
	\end{equation}
	with
	\begin{equation}\label{radial_rescaling_params}
	k_1=-\dfrac{2\iota \omega M}{\eta^2-1}, \qquad k_2=s+\dfrac{\iota \omega M}{\eta\left(\eta+1\right)} \quad \mbox{and} \quad k_3=-1+\dfrac{\iota \omega M}{\eta\left(\eta-1\right)}
	\end{equation}
	in terms of the new variable
	\begin{equation}\label{radial_z}
	z=\dfrac{r}{2M}\dfrac{1-2AM}{1-Ar},
	\end{equation}
	the Eq.~(\ref{radial_eqn}) takes the form of the of general Heun equation~(\ref{heuneqn}) in $u(z)$, with parameters
	\begin{eqnarray}\label{radial_heun_parameters}
	\nonumber 2 \eta a_r=\eta-1, \qquad 2 \eta q_r=\eta\left(1-s^2\right)+E, \hspace{4.8cm} \\
	\alpha_r=1+s, \qquad \beta_r=-\left(1+2k_3\right), \qquad \gamma_r=1-s, \qquad \delta_r=1-s+2k_1,
	\end{eqnarray}
	where $E$ is related to the separation constant $\lambda$ as $\lambda^2=E+s^2$. The subscript $r$ of the parameters stands for radial.
	We now proceed to solve the concerned Heun equation using the $su(1,1)$ structure presented above. We note from Eq.~(\ref{radial_heun_parameters}) that $\gamma_r=1-s=1/2$ or 3/2 for $s=1/2$ or -1/2, respectively. Thus, Eq.~(\ref{factor_condn2}) is satisfied, i.e., the singularity at $z=0$ is elementary. For the singularity at $z=\infty$ to be elementary, one requires $|\alpha_r-\beta_r|=|2+s+2k_3|=1/2$. It is observed that choosing $\omega=0$ in Eq.~(\ref{radial_rescaling_params}) leads to this condition. It may be noted that with this choice the singularity at $z=1$ is also elementary, however, the one at $z=a_r$ is not, in general. We discuss below the solutions of the two radial functions ${}_{\pm1/2}\mathcal{R}(r)$.
	\begin{itemize}
		\item $s=1/2$, solutions for ${}_{1/2}\mathcal{R}(r)$:
		
		We have $\gamma_r=1/2$ (section~\ref{gamma_half}). Additionally we have $\delta_r=1/2$, $\alpha_r =3/2$, $\beta_r=1$. Expressing the two exponents of infinity as $\alpha$ and $\alpha+1/2$, we get $\alpha=\beta_r=1$. This value of $\alpha$ does not belong to the set~(\ref{gamma_half_alpha_values}), hence, the concerned Heun equation does not admit any Heun polynomial solution. One can use the standard series solutions of the Heun equation. 
		
		\item $s=-1/2$, solutions for ${}_{-1/2}\mathcal{R}(r)$:
		
		We have $\gamma_r=3/2$ (section~\ref{gamma_three_half}) and additionally we have $\delta_r=3/2$, $\alpha_r =1/2$, $\beta_r=1$. The two exponents of infinity are of the form $\alpha$ and $\alpha+1/2$ with $\alpha=\alpha_r=1/2$. This value of $\alpha$ belongs to the set~(\ref{gamma_three_half_alpha_values}) and corresponds to the singlet solution given by Eq.~(\ref{case2_singlet_soln}). Hence, we have a solution
		\begin{equation}
		u(z)=\dfrac{1}{\sqrt{z}} \quad \mbox{with eigenvalue} \quad q_r=\dfrac{1}{8} \left(1-\dfrac{3}{\eta}\right).
		\end{equation}
		Plugging this Heun singlet into Eq.~(\ref{radial_rescaling})
		we obtain the following solution for the radial function:
		\begin{equation}
		{}_{-1/2}\mathcal{R}(r)=\sqrt{\dfrac{2M(r-2M)}{(1-\eta)}} \quad \mbox{with} \quad E=-\dfrac{1}{2}\left(\eta + \dfrac{3}{2}\right).
		\end{equation}
		It may be noted that the above solution involves additional Heun polynomial, obtained using the new $su(1,1)$ structure presented here.
	\end{itemize}
	
	\subsection{The polar equation}\label{angular}
	The polar equation takes the form~\cite{bini_bit_ger}
	\begin{equation}\label{angular_eqn}
	\dfrac{1}{\sin \ \theta}\dfrac{d}{d \theta}\left(\sin \ \theta \dfrac{d {}_{s}\mathcal{S}\left(\theta\right)}{d \theta}\right) + V_{\scriptsize\ \mbox{(ang)}}\left(\theta\right) {}_{s}\mathcal{S}\left(\theta\right)=0
	\end{equation}
	where,
	\begin{eqnarray}\label{angular_pot}
	\nonumber V_{\scriptsize\ \mbox{(ang)}}\left( \theta \right) &=& \dfrac{1+E-s^2}{P(\theta)} - \dfrac{1}{P(\theta)^2} \bigg\{\dfrac{\left[s+\left(m-2sM A\right) \cos \ \theta \right]^2}{\mbox{sin}^2 \ \theta} \bigg. \\ 
	&& \bigg. +\left(m+ sM A\right)^2 + \left(1-s^2\right)\left(1- M A \cos \ \theta\right)^2- M^2A^2 \bigg\}.
	\end{eqnarray}
	The label $s$ in ${}_{s}\mathcal{S}\left(\theta\right)$ once again stands for $\pm 1/2$ correspondng to the two polar functions and $m$ is the azimuthal quantum number. With the transformation of the function ${}_{s}\mathcal{S}$ as
	\begin{equation}\label{angular_rescaling}
	{}_{s}\mathcal{S}(\theta)=y(z)P(\theta)^kz^{k_+}\left(z-1\right)^{k_-}, \qquad 2k=s+1+ \dfrac{2m \eta}{\eta^2-1}, \qquad 2k_{\pm}= s- \dfrac{m \eta}{\eta \pm 1}
	\end{equation}
	in terms of the new variable
	\begin{equation}\label{angular_z}
	z=\cos^2 \dfrac{\theta}{2},
	\end{equation} 
	the Eq.~(\ref{angular_eqn}) takes the form of general Heun equation~(\ref{heuneqn}) in $y(z)$ where,
	\begin{eqnarray}\label{angular_heun_parameters}
	\nonumber 2 \eta a_{\theta}=\eta+1 \qquad 2 \eta q_{\theta} = \left(1+2 \eta\right)s^2+\left(1+3\eta\right)s+ \eta-E, \qquad \quad \\
	\alpha_{\theta}=1+s, \qquad \beta_{\theta}=1+2s, \qquad \gamma_{\theta}=1+2k_+, \qquad \delta_{\theta}=1+2k_-.
	\end{eqnarray}
	The subscript $\theta$ of the parameters denotes the polar equation. In order to obtain the solutions of this equation using the methods presented here, we note from Eq.~(\ref{angular_heun_parameters}) that $|\alpha_\theta-\beta_\theta|=|s|=1/2$, i.e., the singularity at $z=\infty$ is elementary. To ensure elementary singularity at $z=0$, one requires $\gamma_\theta=1/2$ or 3/2 (Eq.~(\ref{factor_condn2})). Hence, it is observed from Eq.~\ref{angular_heun_parameters} that one must have $2k_+=\pm 1/2$, which implies (see Eq.~(\ref{angular_rescaling})) that either $m \eta/(\eta + 1)=0$ or $m \eta/(\eta + 1)=\pm1$ (for $s=\pm1/2$, respectively). We note, $0$ is not an allowed value for $m$, since $m$ can assume only half-integer values for spin half particles. Further, the $\eta=0$ limit requires a separate treatment and is beyond the scope of the present work.
	Hence, we stick to $m \eta/(\eta + 1)=\pm1$ and discuss the solutions of the two polar functions ${}_{\pm1/2}\mathcal{S}\left(\theta\right)$ below.
	\begin{itemize}
		\item $s=1/2, \ \mbox{solutions for } {}_{1/2}\mathcal{S}\left(\theta\right)$:
		
		Here we have $m\eta/(\eta+1)=1$, yielding $2k_+=-1/2$, which implies $\gamma_\theta=1/2$ (section~\ref{gamma_half}). In this case the exponents of the singularity at $z=\infty$ are $\alpha_{\theta}=3/2$ and $\beta_{\theta}=2$, i.e., they are of the form $\alpha$ and $\alpha+1/2$ (with $\alpha=\alpha_{\theta}=3/2$). This $\alpha$ value does not belong to the set~(\ref{gamma_half_alpha_values}), which means that the Heun equation in question does not admit Heun polynomial solutions, just as in the case of ${}_{1/2}\mathcal{R}\left(r\right)$. Hence, the solutions to look for are the standard series solutions.
		
		\item $s=-1/2, \ \mbox{solutions for } {}_{-1/2}\mathcal{S}\left(\theta\right)$:
		
		For this case we have $m\eta/(\eta+1)=-1$, giving $2k_+=1/2$ which implies, $\gamma_\theta=3/2$ (section~\ref{gamma_three_half}). Here the allowed $m$ values (leading to $\eta<1$) are $m \in \{-5/2,-7/2,-9/2,...\}$. The singularity exponents at $z=\infty$ are $\alpha_{\theta}=1/2$ and $\beta_{\theta}=0$. Expressing them in the form $\alpha$ and $\alpha+1/2$, we have $\alpha=\beta_{\theta}=0$. This $\alpha$ value belongs to the set~(\ref{gamma_three_half_alpha_values}) and corresponds to the doublet solutions given by Eqs.~(\ref{case2_doublet_soln1}) and (\ref{case2_doublet_soln2}). Hence, we have the solutions
		\begin{eqnarray}
		y_1(z)&=&\dfrac{1}{\sqrt{z}}, \quad \mbox{with eigenvalue} \quad q_{\theta}=-\dfrac{1}{8}\left(3+\dfrac{1}{\eta}\right) \label{ang_heun_soln1} \\
		\mbox{and} \quad y_2(z)&=&\mbox{constant}, \quad \mbox{with eigenvalue} \quad q_{\theta}=0 \label{ang_heun_soln2}
		\end{eqnarray}
		of the Heun equation. Plugging them into Eq.~(\ref{angular_rescaling})
		we obtain the following pair of solutions for the polar function:
		\begin{eqnarray}
		{}_{-1/2}\mathcal{S}\left(\theta\right)&=& %(-1)^{k_-}%
		\dfrac{\left(1-\eta \cos\theta\right)^k\left(\sin\dfrac{\theta}{2}\right)^{2k_-}}{\sqrt{\cos\dfrac{\theta}{2}}} \quad \mbox{with} \quad E=\dfrac{3 \eta}{4} \label{ang_soln_1}\\
		\mbox{and} \hspace{.2cm} {}_{-1/2}\mathcal{S}\left(\theta\right)&=&%(-1)^{k_-}%
		\left(1-\eta \cos\theta\right)^k\sqrt{\cos\dfrac{\theta}{2}}\left(\sin\dfrac{\theta}{2}\right)^{2k_-} \mbox{with} \quad E=-\dfrac{1}{4} \label{ang_soln_2}
		\end{eqnarray}
		where, 
		\begin{equation}\label{angular_params_final}
		k=\left(\dfrac{1}{2}+\dfrac{2}{1-\eta}\right) \qquad \mbox{and} \qquad k_-=-\dfrac{1}{2} \left(\dfrac{1+\eta}{1-\eta}+\dfrac{1}{2}\right).
		\end{equation}
		The above parameter values (Eq.~(\ref{angular_params_final})) are obtained using $m\eta/(\eta+1)=-1$ and $s=-1/2$ in Eq.~(\ref{angular_rescaling}). Among the two solutions above, the one given by Eq.~(\ref{ang_soln_1}) is due to the use of the present $su(1,1)$ generators. The other one can be found using the generators in Ref.~\cite{turbiner} as well.  Thus, the present method has yielded an additional closed form solution for the polar function.
		
		\end{itemize}
	
	\section{Conclusions and outlook}\label{conclusions}
	In this paper we study the dynamics of a massless spin 1/2 particle in the $C$-metric, where the Dirac equation leads to the general Heun equation for the radial and the polar parts. We cast the general Heun operator as a quadratic polynomial of elements of an $su(1,1)$ algebra. This requires the singularities at $z=0$ and $z=\infty$ of the equation to be elementary. Using this new structure, Heun polynomials (including some additional ones that are unavailable from the existent $su(1,1)$ structure~\cite{turbiner}) are obtained from the finite dimensional representation spaces of $su(1,1)$. The general forms of some of the Heun polynomials are listed for convenient use. Some exact solutions for the radial and polar parts of the Dirac equation have been constructed in terms of these Heun polynomials. The obvious future direction of work would be to look for an algebraization of the GHE with general singularity structure, which will yield all the Heun polynomials admitted by the equation and extend it to cover the confluent versions of the Heun equation as well~\cite{my_next_work}.

	\section*{Acknowledgements}
	We thank Prof.~N.~Banerjee and Arunesh~Roy for useful discussions.

\end{document}